\documentclass[usenatbib,fleqn]{mnras}

\usepackage{newtxtext,newtxmath}
\usepackage[T1]{fontenc}
\usepackage{ae,aecompl}

\usepackage{amsmath}	
\usepackage{amssymb}	

\usepackage{ctable}
\usepackage{url}
\usepackage{xspace}
\usepackage[normalem]{ulem}

\def\msun{{\rm\,M_\odot}}

\newcommand{\be}{\begin{equation}}
\newcommand{\ee}{\end{equation}}

\def\h2{${\rm\,H_2}$}

\newcommand{\lir}{$\rm L_{IR}$}

\newcommand\sref[1]{\hyperref[#1]{Section~\ref*{#1}}}
\newcommand\fref[1]{\hyperref[#1]{Fig.~\ref*{#1}}}
\newcommand\Eqref[1]{equation~(\hyperref[#1]{\ref*{#1}})}
\newcommand\tref[1]{\hyperref[#1]{Table~\ref*{#1}}}

\title[Is a top-heavy IMF needed to reproduce SMG number counts?]{Is a top-heavy initial mass function needed to reproduce the submillimetre galaxy number counts?}
\author[Safarzadeh, Lu, \& Hayward]{
	\parbox[t]{\textwidth}{
	Mohammadtaher Safarzadeh$^1$\thanks{E-mail: mts@asu.edu}, Yu Lu$^2$, \& Christopher C.~Hayward$^{3,4}$
	} \vspace*{6pt} \\
	$^1$School of Earth and Space Exploration, Arizona State University, Tempe, AZ 85287-1404, USA\\
	$^2$The Observatories, The Carnegie Institution for Science, 813 Santa Barbara Street, Pasadena, CA 91101, USA\\
	$^3$Center for Computational Astrophysics, Flatiron Institute, 162 Fifth Avenue, New York, NY 10010, USA\\
        $^4$Harvard-Smithsonian Center for Astrophysics, 60 Garden Street, Cambridge, MA 02138, USA
}

\usepackage[all]{hypcap}

\pubyear{2017}

\begin{document}
\label{firstpage}
\pagerange{\pageref{firstpage}--\pageref{lastpage}}
\maketitle

\begin{abstract} 
Matching the number counts and redshift distribution of submillimeter galaxies (SMGs) without invoking modifications to the initial mass function (IMF) has proved challenging
for semi-analytic models (SAMs) of galaxy formation.
We adopt a previously developed SAM that is constrained to match the $z = 0$ galaxy stellar mass function and makes various predictions that agree well with observational constraints; we do not recalibrate the SAM for this work. We implement three prescriptions
to predict the submillimeter flux densities of the model galaxies; two depend solely on star formation rate, whereas the other also depends on the dust mass.
By comparing the predictions of the models, we find that taking into account the dust mass, which affects the dust temperature and thus influences the far-infrared spectral energy
distribution, is crucial for matching the number counts and redshift distribution of SMGs.
Moreover, despite using a standard IMF, our model can match the observed SMG number counts and redshift distribution reasonably well, which contradicts the conclusions of some previous studies that a top-heavy IMF, in addition to taking into account the effect of dust mass, is needed to match these observations.
Although we have not identified the key ingredient that is responsible for our model matching the observed SMG number counts and redshift distribution without IMF variation -- which is challenging given the different prescriptions for physical processes employed in the SAMs of interest -- our results demonstrate that in SAMs, IMF variation is degenerate with other physical processes, such as stellar feedback.

\end{abstract}



\section{Introduction} \label{S:intro}

Submillimeter galaxies \citep[SMGs;][]{Smail:1997,Blain:2002} are rapidly star-forming \citep[$\sim10^2-10^3 \msun~{\rm yr}^{-1}$;][]{Michalowski:2010masses,2012Michalowski}
galaxies located primarily at $z\sim2-4$ \citep[e.g.][]{Chapman:2005,Wardlow:2011,Weiss2013,Simpson2014}
that are believed to be a crucial phase in the formation of the most-massive present-day galaxies. Despite being the subject of extensive study, the nature of SMGs is still debated.

Reproducing the number density and redshift distribution of SMGs has traditionally been challenging for semi-analytic models (SAMs) of galaxy formation.
For example, by coupling the {\sc galform} SAM \citep{Cole:2000} with the {\sc grasil} radiative transfer code \citep{Silva:1998} and therefore self-consistently modeling
dust absorption and emission, \citet{Granato:2000} predicted SMG number counts that were a factor of $\sim20$ less than those observed \citep{Baugh:2005}.
The \citet{Fontanot:2007} model produces an overabundance of bright galaxies at $z < 1$ and yields an SMG redshift distribution that peaks at lower redshift than observed. 
The model of \citet{Somerville:2012} under-predicts the submillimeter (submm) counts by almost two orders of magnitude.

\citet[][hereafter B05]{Baugh:2005} presented an updated version of the {\sc galform} model that was able to match both the observed SMG number counts and redshift distribution
reasonably well. The key change that enabled matching the aforementioned observations was that B05 adopted a flat initial mass function (IMF) in starbursts, which
dominate the SMG population in their model. However, this suggestion was met with significant resistance because
a flat IMF ($dN/d\log M = $ constant) differs drastically from standard IMFs \citep{Kroupa:2001,Chabrier:2003}, which have $dN/d \log M \propto M^{-1.3}$ at the high-mass end
and $M^{-0.3}$ at the low-mass end, and there is no compelling evidence for IMF variation \citep{Bastian:2010}, except perhaps in the centers of massive ellipticals
\citep[e.g.][]{Conroy:2012}, where the IMF is believed to be \emph{bottom-heavy} rather than top-heavy, and the Galactic Center, where a top-heavy IMF ($dN/d \log M
\propto M^{-0.55}$) has been claimed \citep{Bartko2010}.

In the most recent version of the {\sc galform} model \citep[][hereafter L16]{Lacey2015}, the IMF assumed in starbursts is significantly less top-heavy $(dN/d \log M \propto M^{-1}$)
than the flat IMF assumed in starbursts in the B05 model.
However, if a \citet{Kennicutt:1983} IMF is used, the L16 model under-predicts the SMG number counts by as much as two orders of magnitude (see their Fig. C21).
It is thus clear that IMF variation is still a crucial ingredient of their model, and this suggestion remains controversial.

SAMs treat many physical processes, such as star formation and stellar feedback, using prescriptions that can differ considerably amongst SAMs even though they all
are tuned to match key observables, such as the $z = 0$ stellar mass function, well. For this reason, that the aforementioned SAMs were unable to reproduce
the number counts and redshift distribution of SMGs does not preclude alternate SAMs from being able to do so. Moreover, because B05
and L16 tuned their models `by hand' rather than exploring the model parameter space systematically using e.g. Markov Chain Monte Carlo (MCMC)
sampling \citep[e.g.][]{Lu:2011a,Henriques2013,Benson2014}, it is possible that their model could match the observations equally well (or better) by employing a standard IMF
and a different set of parameters.

For the above reasons, in this work, we revisit the question of whether SAMs can reproduce the number counts and redshift distribution of SMGs without
resorting to IMF variation. We employ the SAM of \citet{Lu:2011a}, which is constrained to match observations such as the $z = 0$ stellar mass
function; the optimal parameters are selected via MCMC sampling. A standard \citet{Chabrier:2003} IMF is assumed. To predict the submm flux densities of
the model galaxies, we employ three simple methods from the literature:
a relation from \citet[][hereafter C17]{2017Cowie} that gives the submm flux density as a function of star formation rate (SFR);
the \citet[][hereafter CE01]{Chary:2001} spectral energy distribution (SED) templates, which depend solely on the IR luminosity ($L_{\rm IR}$, which we assume linearly depends on the SFR);
and a fitting function derived from the results of performing dust radiative transfer on hydrodynamical simulations that depends on both the luminosity absorbed by dust and the dust mass \citep{H11,HN13}. We find that as long as we include
the influence of the dust mass on the submm flux density -- namely, that at fixed $L_{\rm IR}$, higher dust mass leads to a colder IR SED and thus
higher submm flux density \citep{H11,Lanz2014,Safarzadeh2016} -- our model matches the observed SMG number counts and redshift distribution reasonably well.

In \sref{S:methods}, we detail our methods. In \sref{S:results}, we present our results. In \sref{S:summary}, we summarize and discuss some implications of our results. 

\section{methods} \label{S:methods}

\subsection{The SAM}
In this paper, we adopt an existing SAM developed in \citet{Lu:2011a, Lu:2014cy}. 
Similar to many other SAMs, this model follows dark matter halo assembly histories and treats various important baryonic processes for galaxy formation, including reionization, radiative cooling, star formation and feedback, and galaxy-galaxy mergers. It predicts the stellar masses, cold gas masses, star formation rates, outflow rates, and metallicities of model galaxies, among other properties.
To address various uncertainties in our understanding of galaxy formation processes, the SAM employs flexible parameterizations for these processes and follows the Bayesian formalism to constrain the model parameters using observational data.
Since we adopt the exact model published in \citet{Lu:2014cy}, i.e. without recalibrating the model in any way, we do not repeat the detailed descriptions of the model in this paper or present a wide variety of predictions but rather refer readers to \citet{Lu:2014cy} for details of the model prescriptions and predictions. 

In summary, the model parameters governing star formation and feedback are tuned using MCMC optimization to match the local galaxy stellar mass function \citep{Moustakas2013}.
The locally constrained model performs well in predicting galaxy properties at high redshift. 
\citet{Lu:2014cy} found that the posterior predictions of the model for the stellar mass function, SFR as a function of galaxy mass, and galaxy gas fractions at a wide range of redshifts (out to $z\sim6$) generally agree with those of the \citet{Croton:2006} and \citet{Somerville:2012} models.
\citet{Lu:2014cy} show that the model achieves remarkable agreement with the observed cosmic star formation rate and the cosmic stellar mass density as functions of redshift to $z\sim 6$.  
Moreover, the model predictions for the gas-phase metallicity match current observational constraints within the observational uncertainties out to $z\sim 2$ \citep[e.g.][]{Kewley2008a, Zahid2014a}.

The model is applied to a set of halo merger trees extracted from a large cosmological $N$-body simulation, the \emph{Bolshoi Planck} simulation \citep{Rodriguez-Puebla2016a} with the cosmology favored by the Planck data \citep{Planck2016}, with parameters $\Omega_{\rm m,0} = 0.307$, $\Omega_{\Lambda,0} = 0.693$, $\Omega_{\rm b,0} = 0.048$, $h = 0.678$, $n = 0.96$, and $\sigma_8 = 0.82$.  
The mass resolution of the simulation allows the model to track halos and subhalos with mass $\gtrsim 7 \times 10^9 \,h^{-1} M_{\odot}$. 
A lightcone with a projected size of $\rm \sim 0.5~deg^{2}$ is used to construct our mock SMG catalogue. We have confirmed that our conclusions are insensitive to cosmic variance.


\subsection{Predicting submm fluxes of the model galaxies}

The model presented in \citet{Lu:2014cy} does not predict SEDs of the simulated galaxies, so for the purposes of this work, we must assign submm flux densities to the model galaxies using the physical properties predicted by the \citet{Lu:2014cy} model without modification.
We employ three methods for this purpose. First, we use the
following relation from C17:
\be
S_{850}= \left(\frac{\rm SFR}{143~\msun~{\rm yr}^{-1} } \right){\rm mJy}.
\ee
We note that this relation depends on SFR alone and thus does not incorporate the influence of the dust mass on the
IR SED shape.

The second method employs the CE01 SED templates. We first calculate $L_{\rm IR}$ using the following conversion \citep{Murphy:2011,Kennicutt:2012}:
\be
{\rm L_{IR}} = 2.6 \times 10^{43} \left(\frac{\rm SFR}{{\rm M}_{\odot}~{\rm yr}^{-1}}\right) {\rm erg~s^{-1}} \,.
\ee
The CE01 templates are parametrized according to their $L_{\rm IR}$; we select the template with the value closest to that computed based on the galaxy's
SFR and then renormalize the template to have the correct $L_{\rm IR}$. We then redshift the SED and convolve it with the SCUBA-2 filter response curve
to calculate $S_{850}$. As with the C17 relation, the submm flux density predicted using this method depends only on the SFR.

Both of the above are purely empirically based relations; we employ them because templates are used in other SAMs \citep[e.g.][]{Somerville:2012} and because they clearly demonstrate the effects of ignoring the effects of dust. However, we also use a physically motivated relation based on the results of performing dust radiative transfer on hydrodynamical simulations of isolated and merging disc galaxies \citep[][hereafter H13]{HN13}. Briefly, the hydrodynamical simulations, which were performed using {\sc gadget-2} \citep{Springel:2005gadget} include the effects of gravity, hydrodynamics, radiative heating and cooling, star formation, black hole accretion, and stellar and AGN feedback; see H13 and references therein (especially \citealt{Springel:2003,Springel:2005feedback}) for details. A suite of isolated and merging disc galaxies with properties (e.g. masses, gas fractions, and dark matter halo profiles) intended to be representative of massive $z \sim 3$ star-forming disc galaxies.

At various times during the snapshots, dust radiative transfer was performed in post-processing using {\sc sunrise} \citep{Jonsson:2006sunrise,Jonsson:2010sunrise}. {\sc sunrise} computes how emission from stars and AGN in the simulation is absorbed, scattered, and re-emitted by dust, including the effects of dust self-absorption, which can be significant in highly obscured galaxies, such as SMGs. Then, motivated by the scaling for an isothermal optically thin modified blackbody, the resulting submm fluxes were fit with a double power law of SFR and dust mass. This yielded the following fitting function, which recovers the submm fluxes of the simulated galaxies with a scatter of $\sim 0.15$ dex:
\be
S_{850}=0.81~{\rm mJy} ~C(z) \left(\frac{\rm SFR}{100~\msun~{\rm yr}^{-1} }\right)^{0.43}  \left(\frac{M_{\rm dust}}{10^8~{\rm M}_{\odot}}\right)^{0.54}\,,
\ee
where
\be
C(z) = \left(\frac{1+z}{3}\right)^{\beta - 1} \left(\frac{D_A(z = 2)}{D_A(z)}\right)^2,
\ee
with $D_A(z)$ denoting the angular diameter distance to redshift $z$,
is a correction factor that accounts for the redshift dependence of the submm flux density (see equation A9 of \citealt{H11}).\footnote{This dependence was not considered in H13
because owing to the negative $k$-correction, it is only relevant for $z \lesssim 0.5$, and galaxies
at those redshifts contributed negligibly to the bright SMG population in H13.} We assume $\beta = 2$.
We note that this relation is accurate despite employing only global properties and neglecting geometry because $L_{\rm IR}$, which is closely tied to the SFR for actively star-forming galaxies \citep{Hayward2014}, and dust mass are the key parameters that determine the far-IR SEDs of galaxies \citep{Safarzadeh2016,Kirkpatrick2017}. Also note that the scaling differs from that expected for an isothermal optically thin modified blackbody because the simulated galaxies -- like real galaxies -- feature a distribution of dust temperatures \citep{H11,Lanz2014}.
Finally, it is worth commenting that this scaling relation also recovers the results of {\sc grasil} \citep{Silva:1998}, which employs simplified axisymmetric geometries to do radiative transfer calculations and is used in the {\sc galform} model (A.~Benson, private communication).

To predict the dust mass of a galaxy, we assume that the dust mass is proportional to the total cold gas mass, $M_{\rm cold}$, and the gas-phase metallicity, $Z_{\rm cold}$, 
\begin{equation}
M_{\rm dust} =  f_{\rm dust} Z_{\rm cold} M_{\rm cold}\,,
\end{equation}
where $f_{\rm dust}$ specifies the fraction of metals that are locked into dust. 
We assume $f_{\rm dust}=0.4$ \citep{Dwek:1998}.


\section{results} \label{S:results}

\begin{figure}
\centering
\vskip -0.0cm
\resizebox{3.0in}{!}{\includegraphics[angle=0]{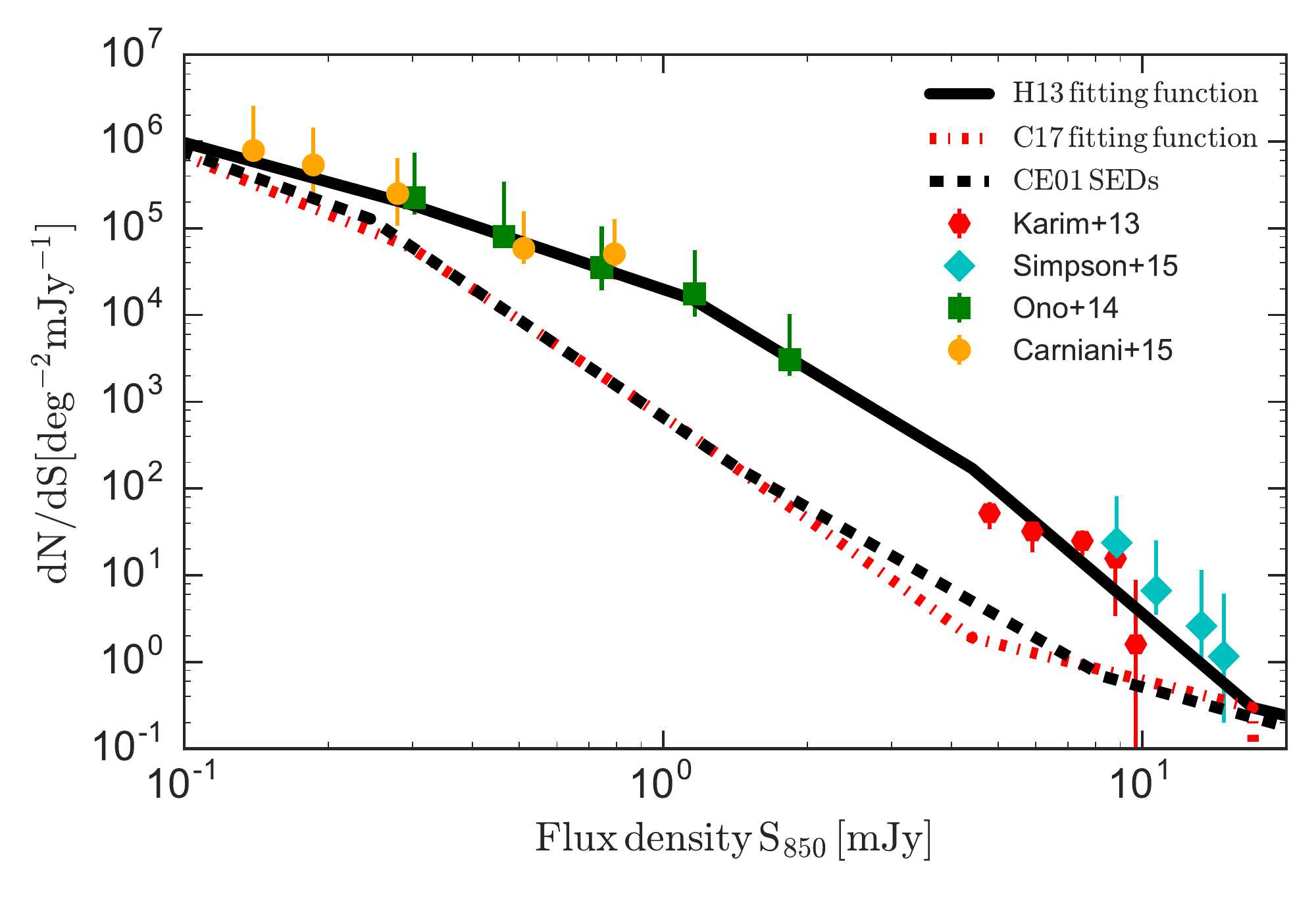}}
\resizebox{3.0in}{!}{\includegraphics[angle=0]{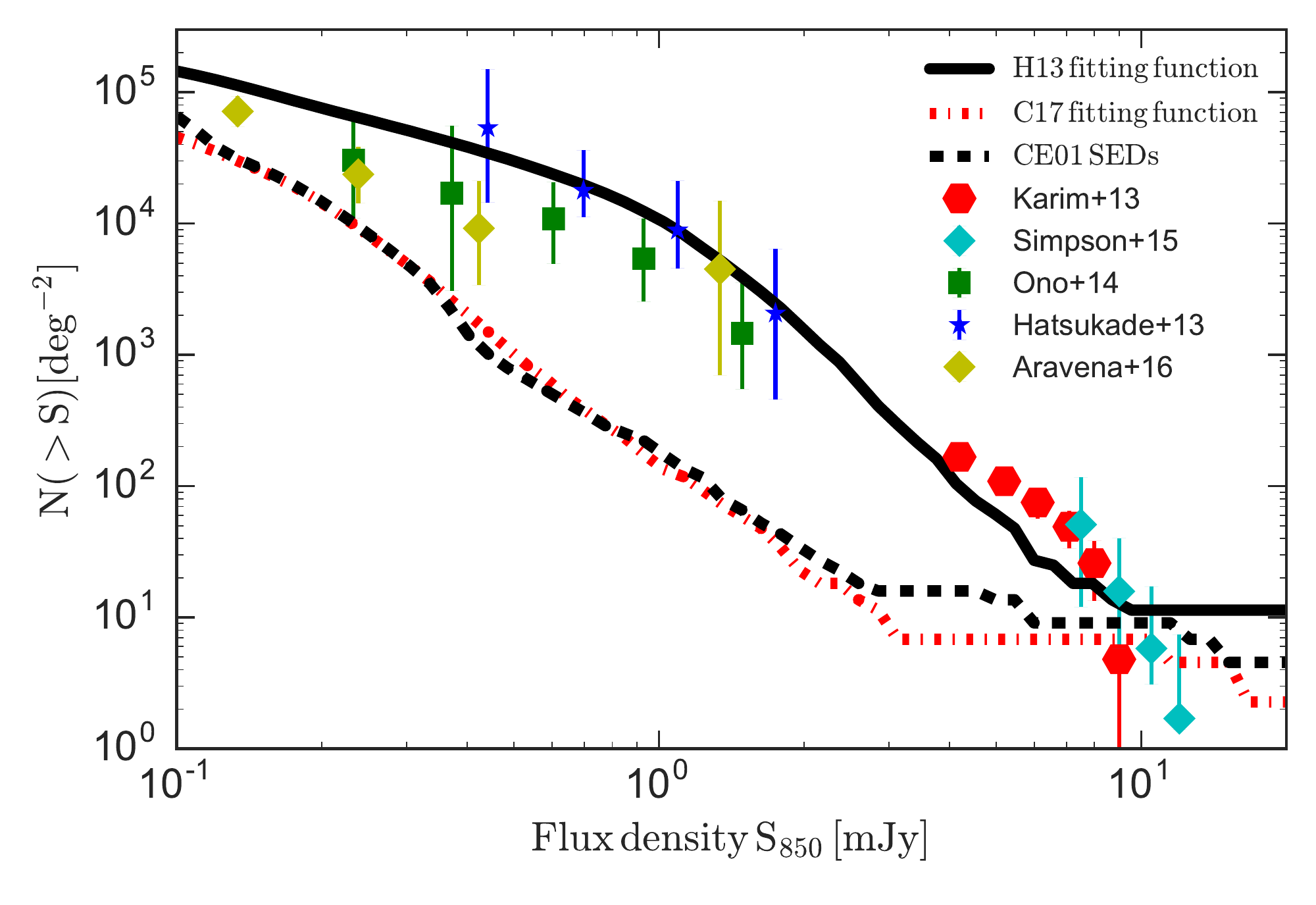}}
\caption{Differential (\emph{top}) and cumulative (\emph{bottom}) 850-\micron~number counts predicted using the C17 SFR-$S_{850}$ relation (\emph{red dash-dotted lines});
the CE01 SED templates, which depend only on $L_{\rm IR}$ (\emph{black dashed lines}); and the $S_{850}(L_{\rm IR}, M_{\rm dust})$ relation from H13 (\emph{black solid lines}).
In both panels, the predictions are compared to counts derived from deep ALMA observations \citep{Karim2013,2013Hatsukade,2014Ono,2015Carniani,2015Simpson,2016Aravena}.
The counts predicted using the C17 relation are very similar to those based on the CE01 templates, and both
are significantly less than those obtained using the H13 fitting function; the latter agree much better with observations, thus demonstrating that accounting for the effects of dust mass is crucial when predicting submm counts.
}
\label{fig:number_count}
\end{figure}

\begin{figure}
\centering
\vskip -0.0cm
\resizebox{3.0in}{!}{\includegraphics[angle=0]{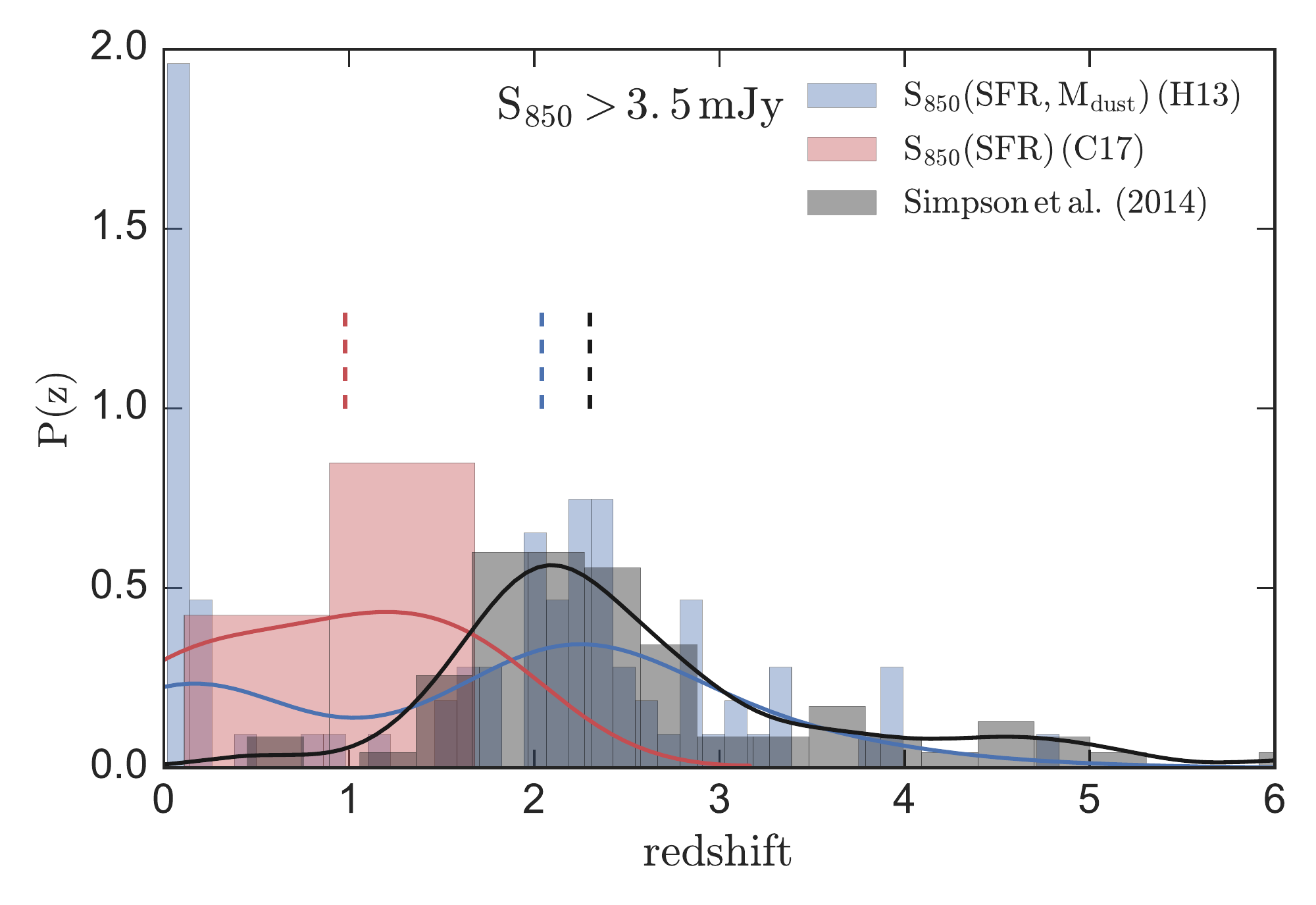}}
\caption{Redshift distributions of model SMGs with $\rm S_{850}>3.5$ mJy predicted using the C17 relation (\emph{red histogram}) and H13 fitting function (\emph{blue}).
The observed redshift distribution of bright SMGs from \citet{Simpson:2014em} is also shown (\emph{grey}). The lines are kernel density estimates to the underlying PDF to guide the eye. The dashed vertical lines denote the median redshifts, $z = 2.2$, 2.0, and 1.0 for the observations, H13 model, and C17 model, respectively. Including the effects of dust mass via the H13 fitting function yields a redshift distribution that peaks at higher redshift and is more consistent with observations compared with the prediction based on the C17 relation, although there is a notable excess of SMGs at $z \lesssim 0.2$ compared with observations.}
\label{fig:redshift_dist}
\end{figure}

\fref{fig:number_count} shows the differential (top panel) and cumulative (bottom panel) number counts predicted from our model compared with recent deep
Atacama Large Millimeter Array (ALMA) observations, which
do not suffer from blending of unresolved sources \citep{Karim2013,2013Hatsukade,2014Ono,2015Carniani,2015Simpson,2016Aravena}.
Consequently, we do not consider the effects of blending, which can cause single-dish submm
number counts to be higher than the true SMG number counts \citep{H12,HN13,HB13}.
In both panels, the prediction based on the C17 SFR-$S_{850}$ relation is shown as a red dash-dotted line,
that based on the CE01 SED templates is indicated by the black dashed line,
and that yielded by the H13 $S_{850}({\rm SFR}, M_{\rm dust})$ relation is shown as a solid black line.
For the observational works in which 1.1-mm counts were reported, we have converted to 850-\micron~counts using $S_{850} \approx 2.3 S_{1.1}$ (H13).
The counts predicted using the three methods differ considerably in the range $S_{850} \sim 0.1-10$ mJy, whereas at lower (not shown) and higher submm fluxes, the number 
counts predicted using the three models are consistent (although as we show below, they do not agree regarding the type of galaxy that dominates a given flux bin).
At $S_{850} \sim 1$ mJy, the counts predicted using the C17 relation or CE01 templates (which are very similar)
are approximately two orders of magnitude less than those obtained using the fitting function.
The counts predicted using the fitting function are qualitatively in much better agreement with the observations, which indicates that incorporating the effects of
dust mass (all else being equal, increased dust mass leads to colder dust and thus higher submm flux density) is crucial for predicting the submm counts (see also B05).

\fref{fig:redshift_dist} shows the redshift distributions of galaxies with $S_{850}> 3.5$ mJy.
The distribution predicted using the C17 relation is shown in red, and that based on the H13 fitting function is shown in blue. The predicted redshift distribution
based on the CE01 templates (not shown) is similar to that obtained with the C17 relation.
The observed distribution from \citet{Simpson:2014em} is shown in grey.
The median redshift for the observed distribution from \citet{Simpson:2014em} is $2.2^{+0.7}_{-0.7}$, whereas the median redshift of $S_{850} > 3.5$ mJy SMGs in the
models based on the H13 fitting function and C17 relation are $2.0^{+0.8}_{-1.9}$ and $1.0^{+0.5}_{-0.6}$, respectively (the errors represent the $16^{\rm th}$ -- $84^{\rm th}$ percentile ranges).
The redshift distribution predicted using the C17 relation (or the CE01 SED templates) peaks at significantly lower redshift than that predicted using the fitting function. 
The redshift distribution predicted using the H13 fitting function agrees reasonably well with the observed distribution from \citet{Simpson:2014em} except that the model features an excess of SMGs at $z \lesssim 0.2$; we speculate about the cause of this discrepancy in next section.

\begin{figure}
\centering
\vskip -0.0cm
\resizebox{3.0in}{!}{\includegraphics[angle=0]{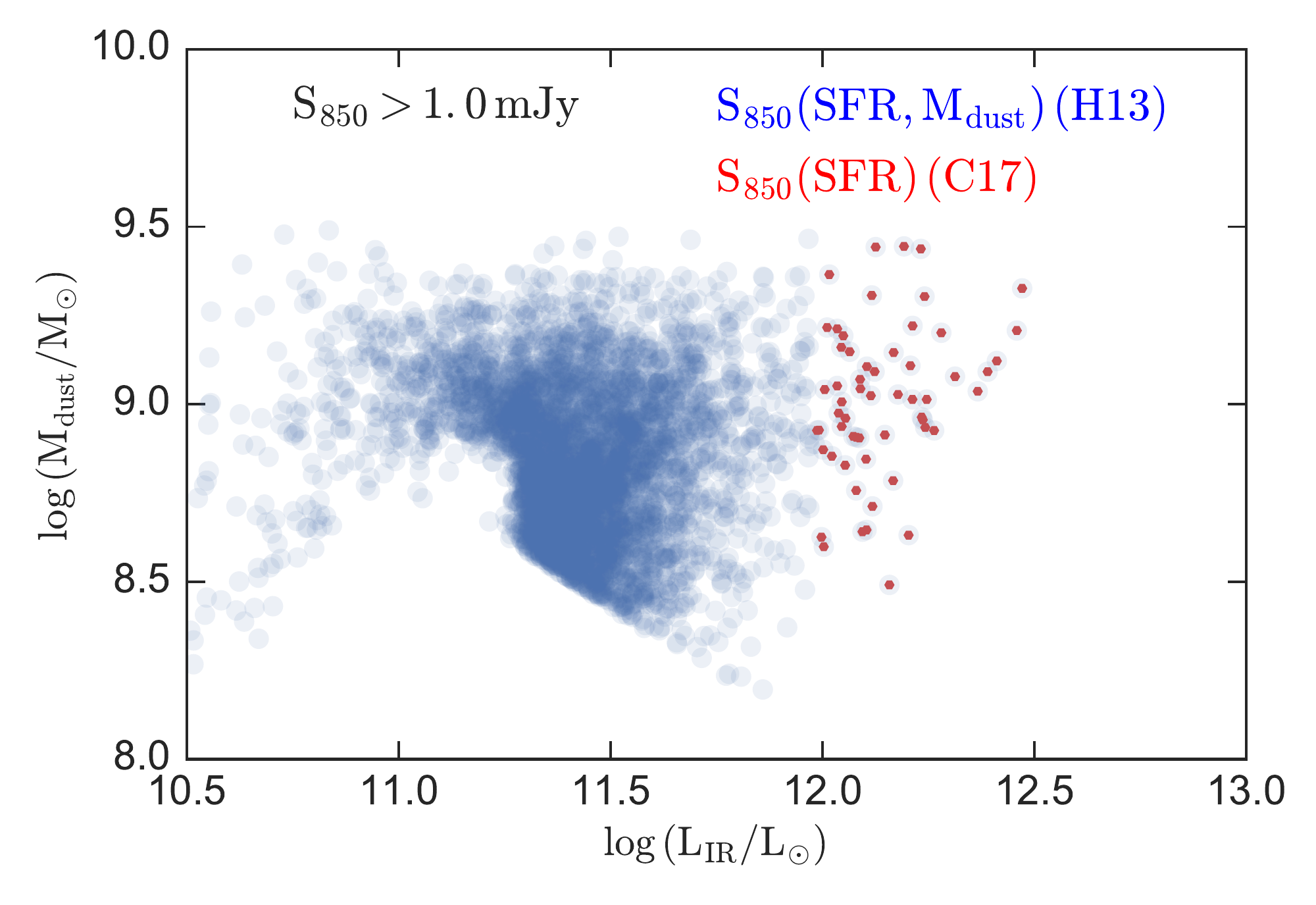}}
\caption{Distributions of the model SMGs in the $\rm M_{dust}$-\lir~plane for the flux range $S_{850} >1$ mJy. The red hexagons correspond to the predictions based on the C17 SFR-$S_{850}$
relation, whereas the blue circles denote the results obtained using the H13 $S_{850}({\rm SFR},M_{\rm dust})$ relation.
When the effects of dust mass are not accounted for (i.e.~the C17 relation is used), the model SMGs tend to have higher $L_{\rm IR}$ and dust mass (the latter holds because
both SFR and dust mass correlate with stellar mass).}

\label{fig:mdust_lir}
\end{figure}

To determine what causes such significant differences between the C17 and H13 predictions, in \fref{fig:mdust_lir}, we plot the distribution of the model SMGs
in the dust mass-$L_{\rm IR}$ plane for the flux bin of $S_{850}> 1$ mJy (we choose this lower flux density cut so that the plane is more densely sampled).
The difference between the two models is clear: the SMGs predicted using the C17 relation (and similarly the CE01 SED templates) tend to have higher $L_{\rm IR}$
and dust mass (the latter holds because in the model, SFR and dust mass both correlate with stellar mass) compared with those predicted using the H13 relation.

\section{summary and discussion} \label{S:summary}

In this work, we have predicted the submm flux of galaxies in the \citet{Lu:2011a} SAM using three different methods.
The first two (C17 and CE01) determine the submm flux density based solely on the SFR.
The second (H13) employs a fitting function derived from performing dust radiative transfer on hydrodynamical simulations; it predicts the
submm flux density given the total SFR and dust mass of a galaxy. 
All models assume a \citet{Chabrier:2003} IMF for star formation and feedback processes regardless of redshift, SFR, or mode of star formation.
We have compared the number counts and redshift distributions of SMGs predicted using the three approaches; those predicted with
the H13 fitting function, which accounts for the effects of variations in dust mass, agree with observations reasonably well, whereas
the C17 relation or CE01 templates, both of which depend only on SFR, yield number counts that are too low and
redshift distributions that peak at significantly lower redshift than observed. 

Above, we noted that when the H13 relation is employed, although the median redshift agrees well with observations, there is an excess of model
SMGs at $z \lesssim 0.2$; it is worth considering how this could be resolved.
First, we note that we have assumed that all star formation is obscured; whereas this is reasonable for high-$z$ galaxies with
SFRs in excess of 100 $\msun~{\rm yr}^{-1}$, this assumption may break down at lower SFR or/and redshift \citep[e.g.][]{Dunlop2017}.
Future work will explore methods for computing the obscured SFR in the SAM.
Second, unlike e.g. B05 and L16,
\emph{we have not tuned} our SAM to match the SMG number counts or redshift distribution; incorporating these into the observations used to constrain the
model parameters and re-doing the MCMC sampling may lead to better agreement between the model and observations, even without changing the physical ingredients of the model.
Regardless of these discrepancies, we note that to the best of our knowledge, of SAMs that have predicted SMG number counts
using a standard IMF, our predictions are in closest agreement with the observed SMG number counts and redshift distribution.

In contrast with previous SAMs, we can reproduce the number counts
and redshift distribution of SMGs -- in addition to various other observational constraints -- reasonably well without resorting to IMF modification.
It is worthwhile examining why this is the case.
First, we have demonstrated that it is important to account for the effect of dust mass in addition to $L_{\rm IR}$ when predicting the submm flux of a model galaxy
(see also B05).
In our model, we find that the submm flux of luminous SMGs can be enhanced by approximately one order of magnitude when the dust mass is taken into account.  
This finding may explain the deficit of luminous SMGs in the study of \citet{Somerville:2012}, who employed SED templates that depend on $L_{\rm IR}$ alone
to predict submm flux densities.

However, the effect of dust mass alone cannot explain the difference between our results and those of e.g. B05 and L16, as
both of these SAMs incorporate the effects of dust mass.
Thus, the differences between the aforementioned two models and ours are likely due to the different prescriptions for physical processes, such as star formation and feedback,
in the models. 
As examples, we highlight two important differences between our model and the {\sc galform} model of B05 that can plausibly significantly affect the predicted submm flux
densities of model galaxies, although there are various other differences between the models that may be as important as or even more important than the two identified here.

First, B05 assume that the efficiency for star formation in the ``quiescent'' mode is proportional to $\sim V_{\rm c}^{-3}$, where $V_{\rm c}$ is the halo circular velocity.
In our model, although we also implement a $V_{\rm c}$ dependence for the star formation efficiency, the power-law index has a much smaller value ($\sim 1$). 
Thus, the star formation efficiency in our model does not depend as strongly on circular velocity (and thus mass and redshift) as in the B05 model. 

Second, our model and that of B05 adopt very different prescriptions for how the outflow mass-loading factor depends on galaxy properties. 
In our model, the outflow mass-loading factor drops rapidly with increasing halo circular velocity ($\eta \propto V_{\rm c}^{-6}$).
For illustration, in our model, a galaxy hosted by a $10^{12} ~\msun$ halo at $z=2$ typically has a star formation rate of $20\msun{\rm yr}^{-1}$ and an outflow rate of $\sim1 ~\msun~{\rm yr}^{-1}$ \citep{Lu:2014cy}, thus allowing high-mass galaxies to retain much of their gas and metals \citep{Lu2015} and consequently have high dust masses. 
In the B05 model, the same galaxy would produce a much stronger outflow: their model adopts a much shallower power law scaling for the outflow mass-loading factor, $\eta=\left({V_c / 150 ~{\rm km\,s^{-1}}}\right)^{-2}$, which yields a mass outflow rate of
$\sim10 ~\msun~{\rm yr}^{-1}$ for a galaxy with the same star formation rate and hosted by the same halo at the same redshift.
In addition, B05 also adopt the ``superwind'' model, which is effective at quenching high-mass galaxies by launching powerful outflows that are never recaptured by the halo \citep{Benson:2003}. 
Due to the ejective nature of these outflows, the effective mass outflow rate for the considered galaxy would be $>10 ~\msun~{\rm yr}^{-1}$ and thus at least an order of magnitude greater than in our model. Consequently, SMGs in the B05 model may retain less of their gas and thus dust than in our model.
Given the drastic differences between these models in terms of predicted mass outflow rates, observational constraints on the mass outflow rates of SMGs may help distinguish amongst these and other models.

In the new version of the {\sc galform} model (L16), both the implicit redshift dependence of the star formation efficiency
and the ``superwind'' model have been removed.
The L16 model is able to match the observed SMG number counts and redshift distribution with an IMF that is considerably less top-heavy
than that employed in B05.
We conclude that differences in, e.g., prescriptions for star formation and feedback amongst SAMs (or/and other physical processes)
are degenerate with the effects of IMF variations,
which casts doubt on the argument that the IMF must be top-heavy in SMGs because of differences between the predictions of SAMs and observed SMGs.

\section*{Acknowledgements}
We are thankful to Henry Ferguson and Andrew Benson for valuable discussion and to Rachel Somerville for providing us with the CE01 templates in electronic form. MTS is supported by NASA under theory grant NNX15AK82G. 
The Bolshoi-Planck simulation was performed within the Bolshoi project of the University of California High-Performance AstroComputing Center (UC-HiPACC) on the SuperMUC supercomputer at LRZ using time granted by PRACE. The Flatiron Institute is supported by the Simons Foundation.

\bibliographystyle{mnras}
\bibliography{std_citations.bib,smg.bib,the_entire_lib.bib}

\bsp
\label{lastpage}

\end{document}